\documentclass[%
mathleft,%
]{an}
\usepackage{graphicx}
\usepackage{times}
\usepackage{amssymb}
\usepackage{hyperref}

\newcommand{\aap}{{A\&A}}%

\begin{document}

\Pagespan{1}{}
\Yearpublication{2011}%
\Yearsubmission{2011}%
\Month{1}%
\Volume{999}%
\Issue{92}%

\title{The kinematics of very low mass dwarfs: splinter session summary}

\author{Adam J.\ Burgasser\inst{1}\fnmsep\thanks{Corresponding author:
  \email{aburgasser@ucsd.edu}}
\and  Jacqueline K.\ Faherty\inst{2}
\and  Sarah Schmidt\inst{3}
\and  Andrew A.\ West\inst{4}
\and  Maria Rosa Zapatero Osorio\inst{5}
\and J.\ Sebastian Pineda\inst{6}
\and Ben Burningham\inst{7}
\and C.\ Nicholls\inst{1}
\and Robyn Sanderson\inst{8}
\and Evgenya Shkolnik\inst{9}
\and David Rodriguez\inst{2}
\and Adric Riedel\inst{10}
\and Viki Joergens\inst{11}
}
\titlerunning{VLM Kinematics}
\authorrunning{A.\ J.\ Burgasser et al.}
\institute{
Department of Physics, UC San Diego, MC 0424, 9500 Gilman Drive, La Jolla, CA 92093, USA
\and Universidad de Chile, Cerro Calan Camino El Observatorio \#1515, Las Condes Santiago, Chile
\and Department of Astronomy, University of Washington, P.O. Box 351580, Seattle, WA 98195, USA
\and Department of Astronomy, Boston University, 725 Commonwealth Ave Boston, MA 02215, USA 
\and Centro de Astrobiolog\'{i}a (CSIC-INTA), Ctra.\ Ajalvir km 4, E-28850 Torrej\'{o}n de Ardoz, Madrid, Spain
\and California Institute of Technology, Department of Astronomy, 1200 E. California Ave, Pasadena CA, 91125, USA
\and Centre for Astrophysics Research, Science and Technology Research Institute, University of Hertfordshire, Hatfield AL10 9AB, UK
\and Kapteyn Astronomical Institute, P.O. Box 800, 9700 AV Groningen, The Netherlands
\and Lowell Observatory, 1400 W. Mars Hill Road, Flagstaff, AZ, 86001, USA
\and Georgia State University, Suite 400, 29 Peachtree Center Ave, Atlanta, GA
30302, USA
\and Zentrum f\"{u}r Astronomie Heidelberg, Institut f\"{u}r Theoretische Astrophysik, Albert-Ueberle-Str.\ 2, 69120, Heidelberg, Germany; Max-Planck Institut f\"ur Astronomie, K\"onigstuhl 17, D-69117 Heidelberg, Germany
}

\received{XXXX}
\accepted{XXXX}
\publonline{XXXX}


\abstract{%
Kinematic investigations are being increasingly deployed in studies of the lowest mass stars and brown dwarfs to investigate their origins, characterize their atmospheres, and examine the evolution of their physical parameters.
This article summarizes the contributions made at the Kinematics of Very Low Mass Dwarfs Splinter Session.
Results discussed include
analysis of kinematic distributions of M, L and T dwarfs; theoretical tools for interpreting these distributions;
identifications of very low mass halo dwarfs and wide companions to nearby stars; 
radial velocity variability among young and very cool brown dwarfs; and
the search and identification of M dwarfs in young moving groups.
A summary of discussion points at the conclusion of the Splinter is also presented.
}

\maketitle

\section{Introduction}
Kinematic studies of stellar populations have long been used to investigate the formation, evolution and structure of the Galactic system.  
While such studies have largely focused on main sequence and giant star populations, the very low-mass dwarfs (VLM dwarfs; M $\lesssim$ 0.1~M$_{\odot}$) are a potentially powerful population for kinematic investigations of the Solar Neighborhood and the Galaxy at large.   VLM dwarfs comprise roughly 20--50\% of ``stars'' in the Galaxy, 
and their longevity make them useful in-situ probes of star formation history and chemical enrichment over the lifetime of the Galaxy.  Substellar VLM dwarfs also evolve (cool) over time, and the interplay between thermal and dynamical evolution can provide complementary insight into the Galaxy's low-mass star formation history and brown dwarf evolutionary theory. 
The kinematics of VLM dwarfs can reveal local young moving groups (YMGs;  Zuckerman \& Song 2004),
high proper motion thick disk or halo subdwarfs (L\'{e}pine \& Scholz 2008; Kirkpatrick et al. 2010),
widely-separated VLM companions to nearby stars (Faherty et al.\ 2010), 
and wide VLM binaries (Caballero 2007; Dhital et al.\ 2010).
All of these are benchmark systems for probing multiple star and brown dwarf
formation and the physical properties of low-temperature atmospheres .
Kinematic monitoring can also reveal radial velocity (RV) and astrometric variables,
systems that enable direct mass measurements (Konopacky et al.\ 2010). 
Finally, kinematic distributions of VLM dwarfs provide statistical measures of age (Wielen 1977). When coupled to rotation and activity,
these measurements can trace the statistical evolution of angular momentum evolution and magnetic field generation (West et al.\ 2008; Irwin et al.\ 2011).  
 
Kinematic investigations of VLM dwarfs have been facilitated in recent years by the discovery of large samples ($\gtrsim$~1000) in wide-field infrared (IR) 
imaging surveys;
the availability of IR astrometry spanning decades; and more sensitive moderate- and high-resolution IR spectrographs (e.g., Magellan/FIRE, VLT/FLAMES).
This splinter session highlighted a cross-section of VLM dwarf science enabled by kinematic investigations.  Section~2 summarizes each of the contributed talks, and
Section~3 summarizes the discussion points made at the end of the splinter.  Presentations and accompanying material can be found online at 
\url{http://www.browndwarfs.org/cs17}.

\section{Contributed talks}

\subsection{Kinematics of the nearest late M, L and T dwarfs 
{\it (Jacqueline Faherty)}}

Kinematic studies of the VLM population has improved considerably over the past decade, and in 2004, we initiated our own program,
the Brown Dwarf Kinematics Project (BDKP; Faherty et al.\ 2009, 2010, 2012a) to further increase both astrometric and RV measurements of VLM dwarfs.  
Two significant results from this program to date are:
\begin{itemize}
\item The largest catalog of The identification of a strong correlation between the kinematics and near-IR colors
of VLM dwarfs, with objects having unusually red/blue $J-K$ colors being kinematically younger/older.  This effect is interpreted as gravity-dependence in the opacities of condensate grains and $H_2$ molecules.
\item Demonstration that luminosities and spectral energy distributions are strongly dependent on age, cloud properties, and metallicity.  Specifically,
young VLM dwarfs are underluminous in near-infrared
bands, despite having larger radii.  This indicates a decrease in effective temperature and increase in cloud opacity with decreasing surface gravity at equivalent classifications. This trend is also been seen in the exoplanets around HR~8799 and 2M~1207 (Currie et al.\ 2011; Barman et al.\ 2011).
\end{itemize}
We are currently measuring 3D space motions of juvenile VLM dwarfs to assign them to specific YMGs (Faherty et al.\ 2012b).


\subsection{M dwarf kinematics and the Galactic disk 
{\it (J.\ Sebastian Pineda)}}

We analyzed the kinematics of $\sim26~000$ M dwarfs from the Sloan Digital Sky Survey (SDSS) Data Release 7 (West et al.\ 2011) to characterize the kinematic distribution of M dwarfs within $\sim$1 kpc. Using a probabilistic Markov Chain Monte Carlo analysis, we were able to statistically separate two kinematic components. Modeling each component as a Gaussian, we estimated their means, dispersions and relative normalization as a function of vertical and radial position in galactocentric coordinates.  We found that the kinematics of M dwarfs match previous SDSS results for solar type stars (Bond et al.\ 2010); specifically, the kinematics of {\em chemically} distinguished thin and thick disk G dwarfs match our {\it kinematic} cold and hot components, respectively. The Gaussian approximation remains a useful way to describe the stellar motions of the thin and thick disk; however, our analysis suggests that the thin/thick disk dichotomy may be an oversimplification. Further study is needed to probe the nature of these different stellar populations and their implications for the evolution of the Galaxy. 


\subsection{The UKIDSS late T dwarf sample: space density, kinematics and wide binaries 
{\it (Ben Burningham)}}

UKIDSS survey operations were completed at the end of May this year, and our follow-up of the UKIDSS late-T dwarf sample is nearing completion, with $\sim$170 T dwarfs now confirmed.  Examination of the T6--T9 sample indicates a space density that scales as $dN/dM \propto M^{-\alpha}$ with $-1 < \alpha < 0$, considerably steeper than cluster surveys (Burningham et al.\ 2010).
Roughly 1400~sq.~deg.\ of sky is covered in two epochs by UKIDSS, allowing for proper motion measurement and selection.  With these, we have identified new candidate halo population brown dwarfs, some with extremely blue IR colors (Murray et al.\ 2011).  We also searched for widely-separated brown dwarf  companions to nearby stars, with five reported to date (Burningham et al.\ 2009, 2011; Day-Jones et al.\ 2011; Murray et al.\ 2011; Pinfield et al.\ 2012).
This sample represents a considerable increase in the number of T-type benchmark systems for tests of atmospheric and evolutionary models.

\subsection{Radial velocities of T Dwarfs with FIRE 
{\it (C.\ Nicholls)}}

We measured radial velocities of 50 nearby T dwarfs using the medium-resolution (R $\sim$ 6000) near-infrared FIRE spectrograph. These measurements represent a four-fold increase in T dwarf RVs.  We performed cross correlation with both radial velocity standards and spectral models, focusing on the highly structured red wing of the $J$ band ($1.27\, \mu m$) peak. Combining these RVs with proper motions from the BDKP yields 3D kinematics, enabling us to measure velocity dispersions and associated kinematic ages. We also report the first RV variables among T dwarfs, including the most tightly-bound brown dwarf pair identified to date with a 1--2 day period and moderate eccentricity.
Derived orbital parameters and masses of such systems provide crucial constraints on brown dwarf evolutionary models and multiple formation theories.


\subsection{Tools for studying the six-dimensional phase space distribution of VLM dwarfs 
{\it (Robyn Sanderson)}}

The ever-growing sample of VLM dwarfs with 6D phase space coordinates opens up a new window into the dynamics of this population. Having complete phase-space information allows us to search for dynamically associated groups of stars even after they are well-mixed into the Galactic disk, through the use of integrals of motion or actions (Helmi \& de Zeeuw 2002). With a large enough sample ($\sim$1000 dwarfs), clustering methods like wavelet transforms can pick out groups in velocity (UVW) space (Antoja et al.\ 2008). This is important for studying the birth and evolution of VLM dwarfs. Complicating this picture are the many processes that can transfer stars and brown dwarfs from their birth sites to different heights and radii in the disk, from resonances with the large-scale Galactic potential (Sellwood \& Binney 2002) to scattering from smaller irregularities like molecular clouds (Wielen 1977). Each of these processes has its own unique signature on the integrals of motion and operates on different timescales (Quinn et al.\ 1993; Sales et al.\ 2009; Di Matteo et al.\ 2010), so full phase-space information is essential to disentangle the dynamic history of VLM dwarfs.

\subsection{Identifying VLM members of young nearby moving groups 
{\it (Evgenya Shkolnik)}}

\begin{figure}
\includegraphics[width=0.9\linewidth]{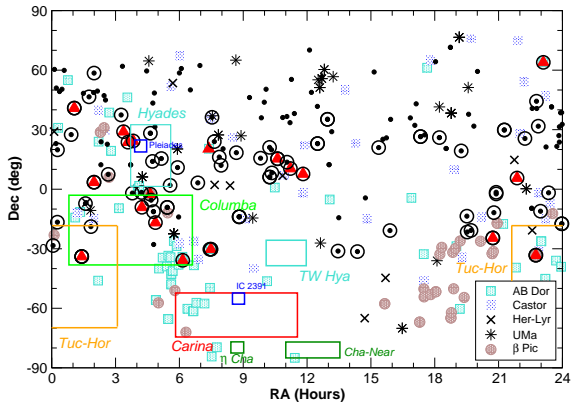}
\caption{Whole sky representation of VLM candidates from Shkolnk et al.\ (2012; black dots) and selected YMGs. Those with
parallaxes are marked with black circles. Stars with a high likelihood of YMG membership  are marked with red triangles.}
\label{fig:shkolnik}
\end{figure}

Young, nearby stars are prime targets for direct imaging searches of extrasolar planets and circumstellar disks. Targets with ages spanning 10 to 100 Myr fill a particularly interesting and relatively unexplored gap, as this time scale coincides with the end of giant planet formation and the beginning of active terrestrial planet formation. These ages are well-sampled by the nearby YMGs, but currently these systems are remarkably poor in M dwarfs. To find the ``missing'' young Ms, we completed a kinematic survey of 165 nearby stars in the field with ages $<$300 Myr (Shkolnik et al.\ 2012). We measured RVs for the complete sample and trigonometric parallaxes for half. Due to their youthful overluminosity and unresolved binarity, the original photometric distances underestimated their true distances by 70\% on average. We identified over a dozen new members of the AB Doradus, Ursa Majoris, Castor, Hercules-Lyra, and $\beta$ Pic YMGs, plus an additional 27 young low-mass stars and 4 brown dwarfs with ages $<$150 Myr which are not associated with any known YMG (Fig.~\ref{fig:shkolnik}). We identified an additional 15 stars which are kinematic matches to one of the YMGs, but whose ages from spectroscopic diagnostics and/or the positions on the sky do not coincide. These results warn against grouping stars based on kinematics alone, and argue that a confluence of evidence is required to claim that a stellar group originates from the same star-forming event.

\subsection{Identifying nearby, young, low-mass stars with the GALEX and WISE catalogs 
{\it (David Rodriguez)}}

The recent release of the all-sky WISE catalog has opened up a new frontier in the search for nearby, young, low-mass stars. Over the last few decades, many $\sim$10--100 Myr-old stars have been identified in YMGs located closer than 100 parsecs of the Earth. Initial searches relied on optical identification (with Tycho and Hipparcos) and X-ray detection (with ROSAT). Recent work has shown that near-IR surveys (like 2MASS) combined with UV data from GALEX can be used to identify additional members in these YMGs (Rodriguez et al.\ 2011; Shkolnik et al.\ 2011). This methodology is well-suited to searching for low-mass stars, which generally lack other membership diagnostics. 
We have now demonstrated that GALEX/WISE color-color diagrams can be used to select young, UV-bright, low-mass stars. Furthermore, the combination of WISE and 2MASS data can be used to estimate proper motions given the 10-year baseline, discriminating young stars from galaxies and old flare stars.  Hundreds of candidate young stars have now been identified via their UV emission and kinematics, and spectroscopic observations will confirm the nature of the candidates. Several show infrared excesses suggesting the presence of warm circumstellar material.


\subsection{Filling in the young M dwarfs
{\it (Adric Riedel})}

Over the past dozen years the RECONS program has obtained photometry,
spectroscopy, and milliarcsecond parallaxes to several known and candidate
young stars.  With the resulting combination of UVW space velocities,
optical excess, spectroscopic surface gravity, H$\alpha$ measurements,
flares, and X-ray emission, we have identified several new young M dwarfs
in the Solar Neighborhood. These include AP Col, the closest pre-main sequence star to the Sun, likely in the 40~Myr Argus/IC 2391 Association (Riedel et al.\ 2011).  We also find evidence that some nearby YMGs may occupy much larger
volumes of space than previously expected.

\subsection{Precision RVs of young brown dwarfs: kinematics, brown dwarf binaries, and their orbits 
{\it (Viki Joergens)}}

\begin{figure}
\includegraphics[width=0.9\linewidth]{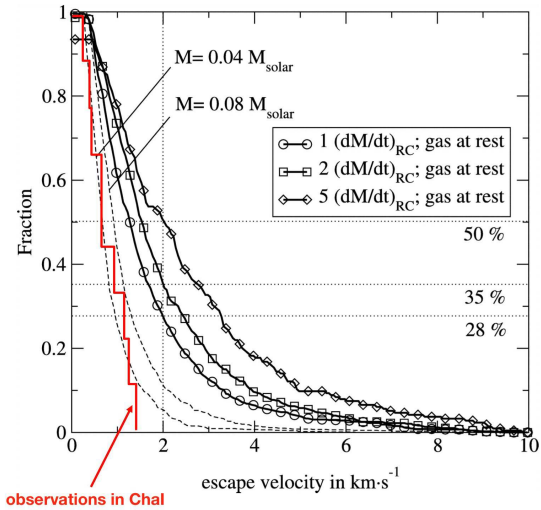}
\caption{Model predictions for the velocities of brown dwarfs formed by the  
ejection scenario in N-body simulations (black lines, Umbreit et al.\ 2005)
compared to observations of VLM dwarfs in Cha I (red line; Joergens 2006).}
\label{fig:vj}
\end{figure}

Precise RVs of very young ($\sim$2\,Myr) VLM dwarfs (spectral types M4-M8)
in Cha~I, based on high-resolution UVES/VLT spectra,
have shown no significant mass dependence or high-velocity  
tail (Joergens 2006; Fig.~\ref{fig:vj}),
as previously predicted by some models of brown dwarf formation 
(Umbreit et al.\ 2005).
RV monitoring of these sources spanning more than a decade has 
led to the detection of  several VLM spectroscopic binaries, including
CHXR\,74 (M4.3; Joergens et al. 2012) and
Cha\,H$\alpha$\,8 (M6), the latter of which contains at least one brown dwarf 
(Joergens et al.\ 2010).
The fraction of very young VLM binaries from this survey, which  
probes separations $\leq$3\,AU, 
is 10$^{+18}_{-8}$\,\% (Joergens 2008), formally
consistent with  values found in star-forming regions from direct
imaging (24$^{+11}_{-7}$\,\% for $\geq$10\,AU; Biller et al.\ 2011) and RV surveys of field VLM dwarfs (3$^{+9}_{-2}$\,\% for $ 
\leq$1\,AU; Blake et al.\ 2010).

\section{Concluding thoughts}

There was considerable interest among participants and audience members
on the identification and characterization of VLM members of YMGs; in particular, the``missing M dwarf'' problem. 
Caution was urged that accurate and complete 3D space motions are 
required to assign membership to 
suspected members of YMGs, with the recognition that as sampling improves the canonical spatial and velocity regions that define these groups may evolve. 
Secondary metrics (surface gravity, Li~I
and activity signatures) are also critical to assigning cluster membership, 
although no one method is necessarily ``best''; each are complementary and together complete the picture of membership.  Additional investigations into convergent point analysis and Bayesian statistical techniques
could provide a more quantitative measure of membership probability.

Radial velocity measurements will increasingly be a crucial component of VLM dwarf kinematics studies as new instrumentation makes such measurements viable.  Such measurements may finally determine the ``true'' multiplicity fraction of VLM dwarfs
through tight constraints on the close binary fraction; and uncover additional systems allowing mass and radius measurements.  Implementation of NIR absorption cells are enabling radial velocity measurements down to 10--100~m/s (Bean et al.\ 2010),
and better use of telluric absorption (in additional to emission) features for RV calibration will  enable searches for companions down to planetary masses.

%
%

\end{document}